\documentclass[12pt]{iopart}

\usepackage[
backend=biber,
style=numeric,
sorting=none
]{biblatex}
\usepackage[colorlinks = true,
            linkcolor = blue,
            urlcolor  = blue,
            citecolor = blue,
            anchorcolor = blue]{hyperref}
            
\AtBeginBibliography{\small}
\usepackage{wrapfig}
         
\usepackage{graphicx}
\usepackage[export]{adjustbox}
\usepackage{etoolbox}
\makeatletter
\patchcmd{\@verbatim}
  {\verbatim@font}
  {\verbatim@font\scriptsize}
  {}{}
\makeatother

\begin{document}

\title[]{A catalog of new Blazar candidates with Open Universe by High School students}

\author{L.~Fronte$^1$, B.~Mazzon$^1$, F.~Metruccio$^1$, N.~Munaretto$^1$ \\
M.~Doro$^{\dagger,2,3}$, P.~Giommi$^{\dagger,4,5}$, I.~Viale$^{2,3}$, U. Barres de Almeida$^6$}

\address{$^1$ Liceo Scientifico Statale U. Morin, via Asseggiano 39, I-30174, Venezia, Italy }
\address{$^2$ University of Padova, Dep. of Physics and Astronomy, via Marzolo 8, I-35131, Padova, Italy}
\address{$^3$ INFN sez. Padova, via Marzolo 8, I-35131, Padova, Italy}
\address{$^4$ Center for Astro, Particle and Planetary Physics (CAP3), New York University Abu Dhabi, PO Box 129188 Abu Dhabi, United Arab Emirates}
\address{$^{5}$ Institute for Advanced Study, Technische Universit{\"a}t M{\"u}nchen, Lichtenbergstrasse 2a, D-85748 Garching bei M\"unchen, Germany}
\address{$^6$ Centro Brasileiro de Pesquisas Fisicas, Rio de Janeiro, Brasil}
\ead{$^\dagger$ michele.doro@unipd.it, giommipaolo@gmail.com}
\vspace{10pt}
\begin{indented}
\item[]September 2022
\end{indented}

\begin{abstract}
Blazars are active galactic nuclei whose ultra-relativistic jets are co-aligned with the observer direction. They emit throughout the whole e.m. spectrum, from radio waves to VHE gamma rays. Not all blazars are discovered. In this work, we propose a catalog of new highly probable candidates based on the association of HE gamma ray emission and radio, X-ray an optical signatures. The relevance of this work is also that it was performed by four high school students from the Liceo Ugo Morin in Venice, Italy using the open-source platform Open Universe in collaboration with the University of Padova. The framework of the activity is the Italian MIUR PCTO programme. The success of this citizen-science experience and results are hereafter reported and discussed.
\end{abstract}

%
\vspace{2pc}

\section{Introduction}
\subsection{The Italian PCTO program}
The Italian Ministry for education, University and Resarch (MIUR, Ministero Istruzione Università e Ricerca) fosters a program called “Percorsi per le Competenze Trasversali e per l’Orientamento” (PCTO)~\cite{pcto}, which stands for “Paths for cross disciplinary skills and orientation”. The program aims at making students acquire skills outside the standard educational program. This may happen outside the school walls. Every student of any school must do a minimum of 90 hours of PCTO experience in order to graduate. The University of Padova has PCTO agreements with several high schools in Italy including the Liceo Scientifico Statale Ugo Morin (LSSUM) in Venice~\cite{lssum}. Within this agreement, a specific program was started with 4 high school students of the 4th year, whose first phase lasted from September 2021 to March 2022, for a total of about 40~h of work for each student, partly at the University premises partly at home.

\subsection{The Open Universe project and the \texttt{Firmamento} portal}
“Open Universe” ~\cite{ou_paper} is an initiative under the auspices of COPUOS (Committee on the Peaceful Uses of Outer Space) with the objective of stimulating a dramatic increase in the availability and usability of space science data, extending the potential of scientific discovery to new participants in all parts of the world and empowering global educational services. Open Universe is now 
defined in detail under the leadership of the United Nations Office Of Outer Space Affairs (UNOOSA). It is officially funded by Brasil and sees collaboration in many participating Countries. The main software infrastructure was a wide-scope Open Universe portal~\cite{ou_portal}, currently not further developed, and a new modern site, specific for blazar science, called \texttt{Firmamento}~\cite{firmamento_portal}  under development at the Center for Astro, Particle and Planetary Physics of the New York University of Abu Dhabi in the United Arab Emirates. \texttt{Firmamento} is also  smartphones friendly\cite{firmamento_paper}. In turns, \texttt{Firmamento} builds on the experience obtained with the tool \texttt{VOU-Blazar}~\cite{VOU-Blazars} in the previous portal~\cite{ou_portal}.

\subsection{Blazars candidates in the Unidentified \texttt{Fermi}-LAT 4FGL DR3 catalog}
For this work the students started with \texttt{Fermi}-LAT data~\cite{fermi}. LAT is a satellite born instrument sensitive to gamma rays in the range $0.1-300$~GeV operating since 2008. In its last 4FGL catalog (DR3)~\cite{4fgl} there are 6658 sources out of which several hundreds are blazars. These are ultrarelativistc jets of particles and radiation with extremely high fluences, formed at super massive black holes when strongly accreting. Blazars are strong emitters at all wavelengths, from radio to very high energies~\cite{agn}.  Out of all unassociated sources in this catalogue, a selection was done based on spectral hardness and distance from the galactic plane. The students were given a list of 198 unidentified LAT sources as input, selected by spectral hardness and location. Their goal was to find counterparts at other wavelengths and eventually propose an identification.

\section{The search for a blazar counterparts}
\begin{figure}[h!t]
    \centering
    \includegraphics[height=6cm,valign=t]{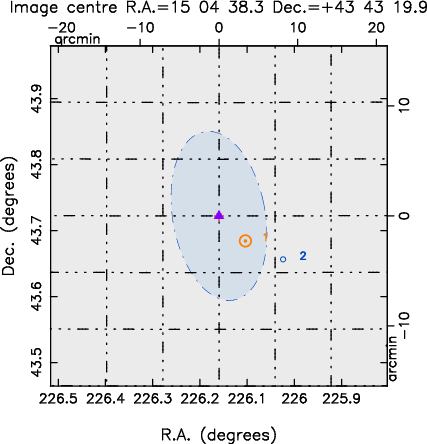}
    \includegraphics[height=5cm,valign=t]{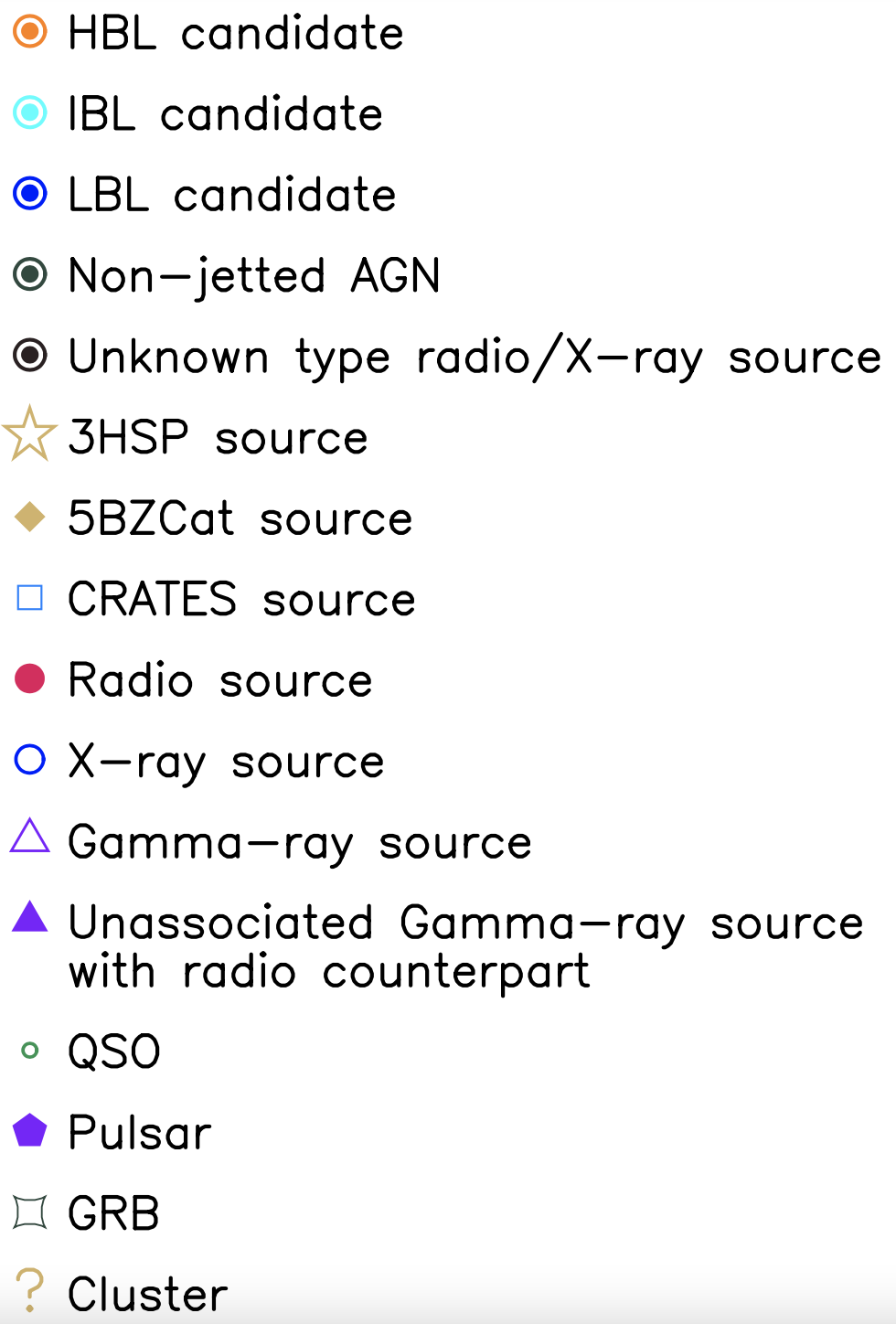}
    \caption{Candidate associations search. The tool find for candidates within the error ellipse of the \textit{Fermi}-LAT candidate (purple triangle) and finds associations that are classified with color-code and size-code markers according to the multi-wavelenghts relations.
    }
    \label{fig:associations}
\end{figure}

The first step was to verify whether these sources had counterparts in any other wavelength. This check was performed with the Open Universe portal~\cite{ou_portal}. The first tool used was \texttt{VOU-Blazars} (now in \texttt{Firmamento}). The instrument takes as input the coordinates of the region and of the uncertainty area that needs to be analyzed and checks for  potential associations in more than 70 different catalogs within this area. The association is based on internal criteria (at the moment not tunable by the user) that weights the existence of counterparts at other wavelengths and rank them according to relative weights. For example, if both an X- and radio counterparts are found at close distance, the significance of the proposed association is ranked high. The results are show in \autoref{fig:associations}.


%

\begin{figure}[h!t]
    \centering
    \includegraphics[height=5cm,valign=t]{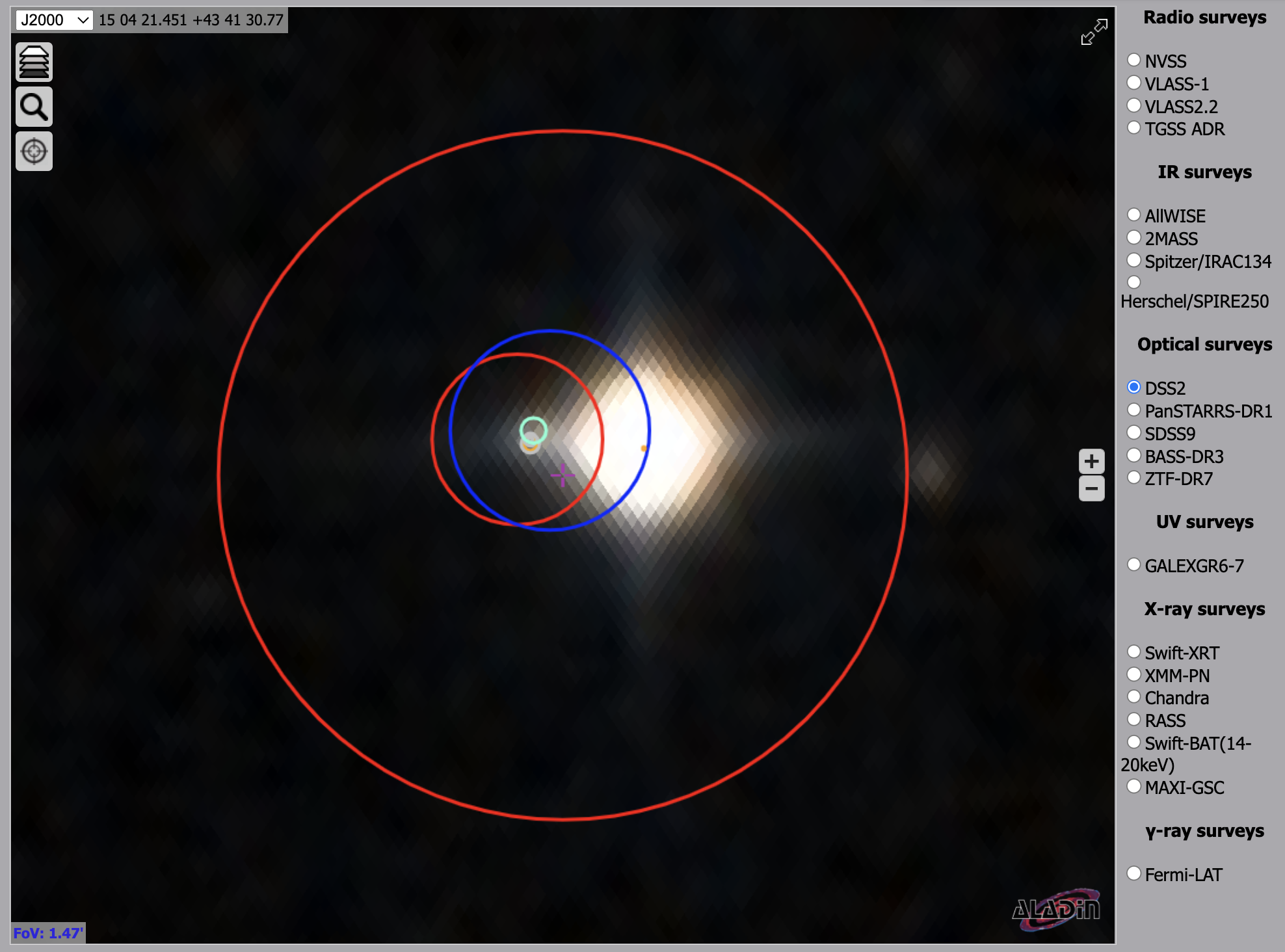}
    \includegraphics[height=7cm,valign=t]{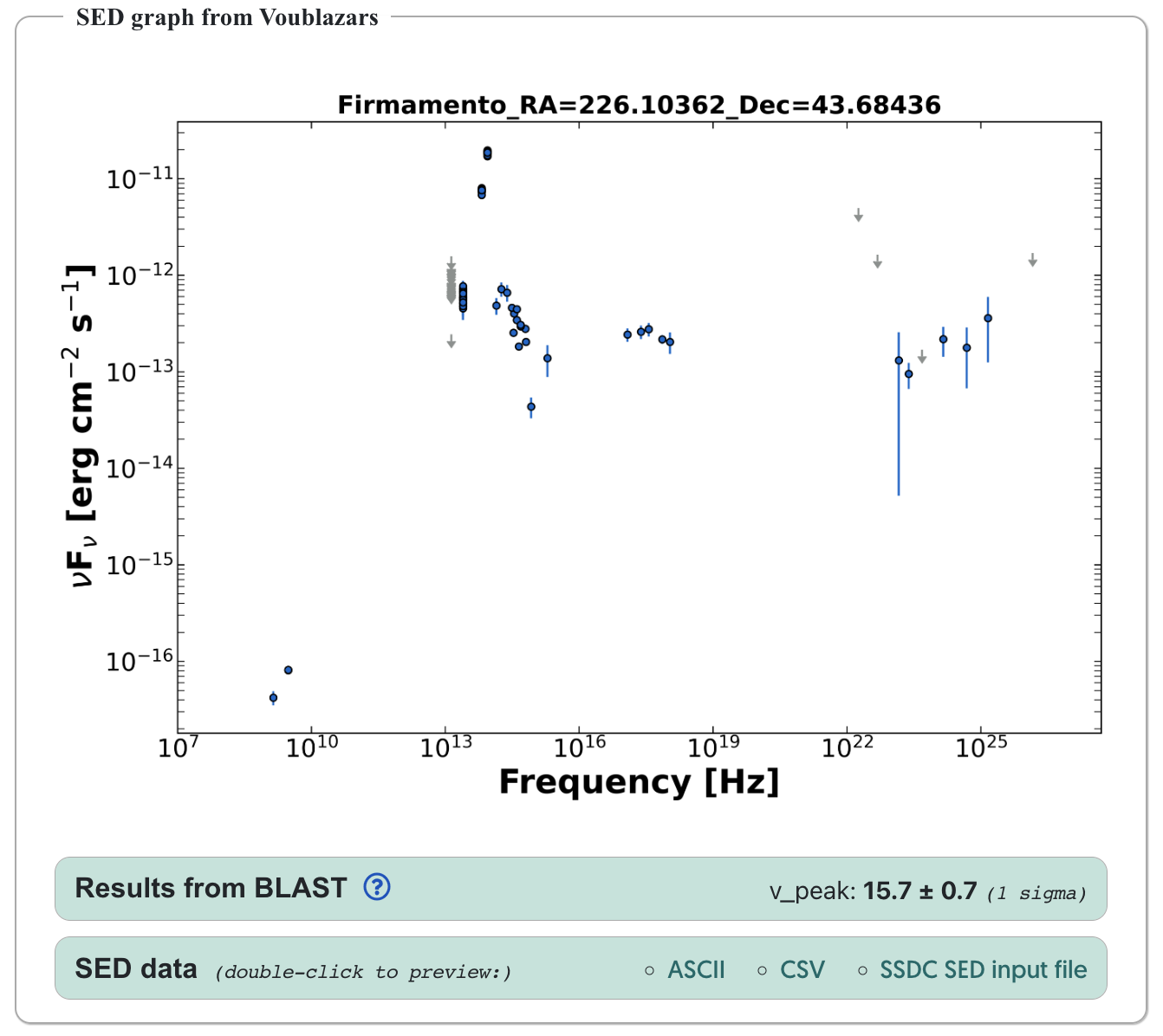}
    \caption{(left) Error circles of the radio (red) and X-ray (blue) catalogue sources around the selected association; (right) The spectral energy distribution of a candidate associated blazar plus the estimated synchrotron peak estimated with \texttt{blast} and the link to download data in table format.}
    \label{fig:sed}
\end{figure}

In a second step, the students verified each single candidate association one by one. This was done by inserting the candidate coordinates in a second tool called \texttt{SSDS Sky survey} (now in \texttt{Firmamento}), also from the Open Universe portal~\cite{ou_portal}. This tool returns a skymap in the optical spectrum around the direction of the candidate, as shown in \autoref{fig:sed} (left). In a third step, the students used the tool called \texttt{VOU-SED} (now in \texttt{Firmamento}), that generates the Spectral Energy Distribution (SED), as shown in \autoref{fig:sed} (right) and a table data file, e.g.:

\begin{center}
    \begin{verbatim}
	   1  matched source    0.11633   25.46800  99
 Frequency     nufnu     nufnu unc.  nufnu unc. start  end   Catalog      Reference
    Hz       erg/cm2/s     upper       lower    MJD    MJD   
------------------------------------------------------------------------------------------------------------
 1.400E+09   1.372E-16   1.498E-16   1.246E-16  55000.  55000.  NVSS      Condon et al. 1998, AJ, 115, 1693             
 2.418E+17   1.185E-12   1.780E-12   5.893E-13  55000.  55000.  XMMSL     Saxton et al. 2008, A&A, 480, 611
 2.660E+17   9.824E-13   1.473E-12   4.916E-13  55000.  55000.  XMMSL     Saxton et al. 2008, A&A, 480, 611
 1.692E+18   0.000E+00   0.000E+00   0.000E+00  55000.  55000.  XMMSL     Saxton et al. 2008, A&A, 480, 611
 4.455E+14   2.257E-13   0.000E+00   0.000E+00  55000.  55000.  GAIA      The Gaia Coll. 2016, A&A, 595, A2
 8.328E+14   0.000E+00   0.000E+00   0.000E+00  55000.  55000.  HST       Lasker et al. 2008, AJ, 136, 735
 6.813E+14   1.881E-13   2.434E-13   1.453E-13  55000.  55000.  HST       Lasker et al. 2008, AJ, 136, 735
 5.451E+14   0.000E+00   0.000E+00   0.000E+00  55000.  55000.  HST       Lasker et al. 2008, AJ, 136, 735
 4.684E+14   0.000E+00   0.000E+00   0.000E+00  55000.  55000.  HST       Lasker et al. 2008, AJ, 136, 735
 3.795E+14   0.000E+00   0.000E+00   0.000E+00  55000.  55000.  HST       Lasker et al. 2008, AJ, 136, 735
 6.233E+14   2.748E-13   2.791E-13   2.706E-13  55000.  55000.  PANSTARRS Chambers et al. 2016 1612.05560
 ...     
\end{verbatim}
\end{center}

The students ultimately applied a program called \texttt{blast} (Blazar Synchrotron Tool)~\cite{blast}, based on  a machine learning algorithm, to estimate the position of the synchrotron peak of a blazar given their SED as a txt file (now automatically computed in \texttt{Firmamento}, see \autoref{fig:sed}). This position is important to classify blazars into sub-categories (e.g. Low-Energy peaked, High-Energy peaked etc.)


\section{The catalog of new blazar candidates}
\label{sec:catalog}
We have collected our results in \autoref{tab:catalog}. The table reports the \texttt{Fermi}-LAT 4FGL name and position on the left hand side and on the right hand side some basic information of the best candidate: the catalog ID name, its position, the redshift when available, and the position and uncertainty of the synchrotron peak computed by the \texttt{blast} code~\cite{blast}. The catalog acronym LSSUM stands for Liceo Scientifico Statale Ugo Morin.

\begin{figure}[h!t]
    \centering
    \includegraphics[width=0.32\linewidth]{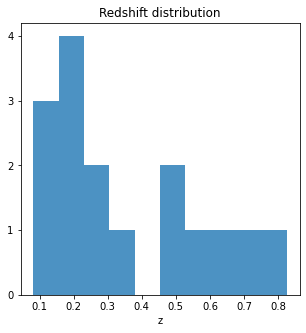}
        \includegraphics[width=0.32\linewidth]{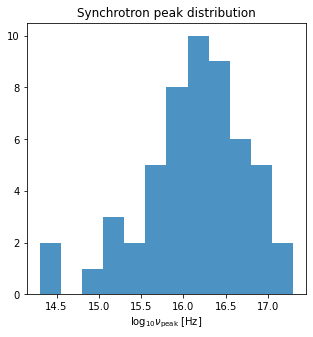}
    \caption{(left) Redshift distribution and (right) Synchrotron Peak Distribution of the LSSUM targets of \autoref{tab:catalog}.}
    \label{fig:my_label}
\end{figure}

In \autoref{fig:my_label} we show the redshift and synchrotron peak distribution of the associations. Both makes sense: the redshift distribution peaks at low distances and the synchrotron peak distribution is somewhat symmetrically peaked around $10^{16}$~Hz, in accord with our cuts in spectral hardness of starting \textit{Fermi}-LAT candidate set.

\begin{table}[h!t]
  \centering
  \begin{scriptsize}
  \begin{tabular}{lcc|lcccc}
  \hline\hline
\texttt{Fermi}-LAT ID                &  RA & Dec & LSSUM ID & RA & Dec & $z$ & $\log_{10}\nu_{\rm{peak}}$ \\
\hline
4FGL J0000.7+2530	& 0.188	& 25.515	& LSSUM J000027.9+252805	& 0.11633& 25.4680	& 0.49	& 16.6	$\pm$0.5\\
4FGL J0026.1$-$0732	&6.540	&-7.543	    & LSSUM J002611.6$-$073115	&6.54842 &-7.52097	& -	&16.9	$\pm$0.4\\
4FGL J0045.8$-$1324	&11.472	&-13.403	& LSSUM J004602.8$-$132422	& 11.51154 & -13.4060&-	&15.6	$\pm$0.6\\
4FGL J0055.7+4507	&13.940	&45.124	    & LSSUM J005542.7+450701	&13.92792	&45.11706 &-	&15.8	$\pm$0.5\\
4FGL J0152.9$-$1109	&28.237	&-11.162	&  LSSUM J015313.2$-$110627 &28.30487	&-11.10753	&-	&16.4	$\pm$0.4\\
4FGL J0159.8$-$2234	&29.951	&-22.576	& LSSUM J015946.9$-$223246	&29.94541 & -22.54637	&-	&16.1	$\pm$0.5\\
4FGL J0231.0+3505	&37.775	&35.100	    & LSSUM J023112.2+350444	& 37.80078 & 35.07916	&-	&17.2	$\pm$0.5\\
4FGL J0249.2+1652	&42.303	&16.882	    & LSSUM J024902.8+165259	& 42.26183 & 16.88331	&-	&16.3	$\pm$0.4\\
4FGL J0251.1$-$1830	&42.784	&-18.509	& LSSUM J025111.5$-$183112	&42.79800 & -18.52019	&-	&15.7	$\pm$0.5\\
4FGL J0357.7$-$6808	&59.440	&-68.134	& LSSUM J035732.10-680932	&59.38733 &-68.15911 & 0.0886 &16.1	$\pm$0.4\\
4FGL J0420.6$-$4802	& 65.173	&-48.048	& LSSUM J042038.7$-$475703	&65.16121 & -47.95103	&-	&16.3	$\pm$0.6\\
4FGL J0438.0$-$7329	&69.524	&-73.485	& LSSUM J043837.1$-$732921	&69.65446 & -73.48933	&0.0948&15.5	$\pm$0.3\\
4FGL J0539.2$-$6333 	&85.055	&6.917	    & LSSUM J053855.9$-$633239	&84.73275 & -63.54431&-	& 14.1	$\pm$1.0 \\
4FGL J0625.5+7029 	&96.392	&70.497	    & LSSUM J062534.7+702943	& 96.39471 & 70.49533 &	-	&16.3	$\pm$0.7\\
4FGL J0641.4+3349	&100.356&	33.820	& LSSUM J064142.0+334928 &100.42501	&33.82450	&-	&14.3	$\pm$0.6\\
4FGL J0751.2$-$0029 	&117.812&	-0.488	& LSSUM J011119.2$-$002748	&17.830002 & -0.463578	&-	&15.7	$\pm$0.4\\
4FGL J0800.9+0733	&120.226&	7.551	& LSSUM J080056.5+073235	&120.23562&	7.54308	&-	&15.5 $\pm$0.4	\\
4FGL J0815.5+6554 	&123.880&	65.900	& LSSUM J081539.8+655004	&123.91592& 65.83461	&-	&16.1	$\pm$0.3\\
4FGL J0838.5+4013	&129.629&	40.224	& LSSUM J083903.1+401545	&129.76283 &	40.26272 &0.194	&15.7	$\pm$0.4\\
4FGL J0903.5+4057	&135.899&	40.962	& LSSUM J090314.7+405559	&135.81129 & 40.93325	&0.188 	&16.2	$\pm$0.5\\
4FGL J0914.5+6845	&138.647&	68.751	& LSSUM J091429.7+684508 &138.62379 &	68.75242	&-	&15.9	$\pm$0.5\\
4FGL J0944.6+5729 	&146.090&	-9.192	& LSSUM J94432.3+573536 &146.13471 &	57.59336	&0.72	&16.1	$\pm$0.4\\
4FGL J1047.2+6740 	&161.820&	67.674	& LSSUM J104705.9+673758	&161.77458 &67.63278	&-	&16.4	$\pm$0.6\\
4FGL J1118.1+5857 	&169.542&	58.965	& LSSUM J111709.8+585921	&169.29063 &58.98917	&0.0814	&15.8	$\pm$0.6\\
4FGL J1146.0$-$0638	&176.502&	-6.638	& LSSUM J114600.8$-$063854 & 176.50354&	-6.64858	&-	&16.1	$\pm$0.4\\
4FGL J1155.2$-$1111	&178.820&	-11.189	& LSSUM J115514.9$-$111122	&178.81192 & -11.18958&	-	&16.6	$\pm$0.5\\
4FGL J1158.8$-$1430	&179.709&	-14.501	& LSSUM J115912.6$-$143154	&179.80263 &-14.53189&	-	&17.0	$\pm$0.5\\
4FGL J1403.7+2429	&210.936&	24.495	& LSSUM J140350.3+243304	&210.95954 & 24.55133	&0.343	&16.2	$\pm$0.4\\
4FGL J1409.8+7921	&212.464	&79.351	& LSSUM J141046.4+792412	&212.69314 & 79.40343	&-	&14.3	$\pm$0.6\\
4FGL J1441.4$-$1934	&220.350&	-19.578	& LSSUM J144127.1$-$193552	&220.36650& -19.59789&	-	&15.7	$\pm$0.5\\
4FGL J1452.0$-$4148	&223.017&	-41.804	& LSSUM J145224.6$-$414948	&223.10243 & -41.82995&	-	&15.2	$\pm$0.6\\
4FGL J1504.6+4343 & 226.159 & 43.722 & LSSUM J150425.1+434106 & 226.10468 &   43.68521 &- &15.7	$\pm$0.7\\
4FGL J1519.7+6727	&229.943&	67.458	& LSSUM J152000.4+672613	&230.00179 & 67.43703	&-	&15.0	$\pm$0.4\\
4FGL J1544.9+3218	&236.239&	32.304	& LSSUM J154433.2+322148	&236.13829 &32.36350	&featureless	&15.2	$\pm$0.5\\
4FGL J1554.2+2008	&238.553&	20.148	& LSSUM J155424.1+201125 &238.60050 & 20.19039	&0.222 &16.9	$\pm$0.8\\
4FGL J1626.5+6257	&246.644&	62.959	& LSSUM J162646.0+630048	&246.69188 &63.01350	&0.24 (Phot.)	&16.8	$\pm$0.5\\
4FGL J1628.2+4642	&247.063&	46.715	& LSSUM J162755+464249	&246.98105 & 46.71342 &0.2135	&15.8	$\pm$0.4\\
4FGL J1658.5+4315	&254.646&	43.254	& LSSUM J165831.5+431615 &254.63126 & 43.27085	&0.63 (Phot)	&16.0	$\pm$0.5\\
4FGL J1706.4+6428	&256.606&	64.475	& LSSUM J170623.3+642725	&256.59688 & 64.45706	&0.27 (Phot) &15.9	$\pm$0.7\\
4FGL J1727.1+5955	&261.776&	59.926	& LSSUM J172640.4+595549	&261.66833 & 59.93036	&featureless&16.1	$\pm$0.4\\
4FGL J1923.0$-$4746	&290.752&	-47.769	& LSSUM J192304.4$-$474501	&290.76829 & -47.75053 &	-	&16.7	$\pm$0.5\\
4FGL J1928.5+5339	&292.139&	53.653	& LSSUM J192833.6+533902	&292.14005 & 53.65058 &-	&17.3	$\pm$0.4\\
4FGL J2012.1$-$5234	&303.039&	-52.570	& LSSUM J201213.7$-$523251	&303.05712 & -52.54753&	-	&16.2	$\pm$0.4\\
4FGL J2020.7$-$4536	&305.198&	-45.614	& LSSUM J202022.9$-$452924	&305.09543 & -45.49008 &	-	&16.6	$\pm$0.5\\
4FGL J2022.3+0413	&305.598&	4.222	& LSSUM J202225.1+041235 &305.60469& 4.209826	&-	&16.7	$\pm$0.6\\
4FGL J2028.8$-$0010	&307.215&	-0.171	& LSSUM J202850.4$-$000840	& 307.20998 &	 -0.14451&-	&16.3	$\pm$0.5\\
4FGL J2030.3$-$5038	&307.590&	-50.634	& LSSUM J203024.0$-$503413	& 307.60017 & -50.57028&	-	&16.6	$\pm$0.4\\
4FGL J2038.7$-$3655	&309.686&	-36.925	& LSSUM J203839.10$-$365426 &309.66664 & -36.90732&	-	&16.1	$\pm$0.6\\
4FGL J2142.5$-$2029	&325.642&	-20.497	& LSSUM J214239.8$-$202819	&325.66575 & -20.47197&	-	&16.4 $\pm$0.4\\
4FGL J2144.8$-$1600	&326.216&	-16.010	& LSSUM J214439.1$-$155931	&326.16300 & -15.99200&	-	&16.4	$\pm$0.6\\
4FGL J2207.1+2222	&331.791&	22.374	& LSSUM J220704.1+222231	&331.76713 & 22.37542	&0.557	&15.8	$\pm$0.5\\
4FGL J2217.0$-$6727	&334.255&	-67.453	& LSSUM J221659.5$-$672800	&334.24813 & -67.46672&	-	&16.4$\pm$0.5\\
4FGL J2237.2$-$6726	&339.304&	-67.437	& LSSUM J223709.4$-$672618	&339.28917 & -67.43861&	-	&15.9$\pm$0.4\\
4FGL J2237.8+2430	&339.458&	24.511	& LSSUM J223738.2+243256	&339.40900 & 24.54910	&0.50	&15.9$\pm$0.4\\
\hline\hline
\end{tabular}
  \end{scriptsize}
\caption{\label{tab:catalog}The preliminary LSSUM (Liceo Scientifico Statale Ugo Morin) catalog of blazar candidates. In the redshift column, "phot" means the redshift is taken from SDSS17 or NED and not from the galaxy spectra, while "featureless" is in case there are no optical lines.}
\end{table}

\section{Outlooks and Conclusions} 

Although we regard our associations as highly reliable, \autoref{tab:catalog} is only preliminary
and incomplete. We plan to re-evaluate the targets by relaxing our criteria on the number of \texttt{Fermi}-LAT unassociated sample. After that, our aim is to try and validate as much as possible our candidates. In some cases, we need additional data. We are discussing the possibility to carry out proposal of observation in optical (for redshift estimation) and in X-ray and gamma-ray to validate the inverse Compton peak. We hope our achievements will end in a journal publication.

\medskip\noindent We close with the students' thoughts:
%
\begin{quote}
\textit{
“This PCTO experience has been a fundamental opportunity to grow as persons, it gave us the possibility to see the research environment in close contact and to understand what working at a University really means. Thanks to this occasion we now know that we’d like to have a career as researchers someday.”}
\end{quote}

\begin{small}
\paragraph{Acknowledgements.} We warmly thank prof. Arianna Boldrin and prof. Alice Scelsi, respectively responsible for the PCTO program at LSSUM and for this specific program with UniPD. We also thank prof. Elisa Prandini for useful discussions about the project and the UNIPD personnel who supported this initiative. 
\end{small}

\printbibliography

\end{document}